\documentstyle[12pt,aaspp4]{article}
\begin{document}

%
%

\title{The Abundance of Mg in the Interstellar Medium \footnote[1]{Based on observations with the NASA/ESA {\em Hubble Space Telescope}, obtained at the Space Telescope Science Institute, which is operated by the Association of Universities for Research in Astronomy, under NASA contract NAS5-2655.}}

\author{Edward L. Fitzpatrick}
\affil{Princeton University Observatory, Peyton Hall, Princeton, New Jersey 08544}

\begin{abstract}

An empirical determination of the $f$-values of the far-UV Mg II
$\lambda\lambda$1239,1240 lines is reported.  The strong near-UV Mg II
$\lambda\lambda$2796,2803 lines are generally highly saturated along
most interstellar sightlines outside the local ISM and usually yield
extremely uncertain estimates of Mg$^+$ column densities in
interstellar gas.  Since Mg$^+$ is the dominant form of Mg in the
neutral ISM and since Mg is expected to be a significant constituent of
interstellar dust grains, the far-UV lines are critical for assessing
the role of this important element in the ISM.  This study consists of
complete component analyses of the absorption along the lines of sight
toward HD 93521 in the Galactic halo and $\xi$ Per and $\zeta$ Oph in
the Galactic disk, including all four UV Mg$^+$ lines and numerous
other transitions.  The three analyses yield consistent determinations
of the $\lambda\lambda$1239,1240 $f$-values, with weighted means of
$6.4\pm0.4\times 10^{-4}$ and $3.2\pm0.2\times 10^{-4}$, respectively.
These results are a factor of $\sim$2.4 larger than a commonly used
theoretical estimate, and a factor of $\sim$2 smaller than a recently
suggested empirical revision.  The effects of this result on gas- and
dust-phase abundance measurements of Mg are discussed.

\end{abstract}


%
%

\section{Background}

The combination of high spectral resolution, photometric precision, and
sensitivity provided by the {\it Goddard High Resolution Spectrograph}
({\it GHRS}) has transformed the study of UV interstellar absorption
lines, enabling detailed examination of individual absorbing regions in
the ISM.  The potential precision of these measurements has also
triggered new interest in the determination of the atomic constants,
particularly the oscillator strengths ($f$-values), required to convert
measured line strengths into gas-phase column densities.  Recent
improvements in the available body of astrophysically interesting
$f$-values have come from empirical studies using interstellar
absorption lines, new laboratory measurements, and theoretical
calculations (see Table 2 of Savage \& Sembach 1996a).

One important species for which the $f$-values remain an impediment to
determining accurate column densities is Mg$^+$, the dominant form of
Mg in H I gas.  Along most interstellar sightlines --- with the
exception of those that pass only through the local ISM (e.g., Linsky
\& Wood 1996) --- the near-UV Mg II $\lambda\lambda$2796,2803 lines are
strongly saturated and yield limited column density information.  The
natural source of accurate Mg$^+$ column densities is thus the only
other pair of observationally accessible Mg$^+$ lines, the intrinsically
much weaker $\lambda\lambda$1239,1240 doublet.  The usefulness of these
lines has been compromised, however, by uncertainty in the $f$-values.
In recent years, the theoretical calculations from Hibbert et al.
(1983) have been used by many investigators.  However, Sofia, Cardelli,
\& Savage (1994) suggested that these values are a factor of $\sim$4.6
too small, based on an empirical investigation of absorption towards
the star $\xi$ Per.  These two results, listed in the first two lines
of Table 1, are mutually inconsistent given their quoted uncertainties
and imply Mg$^+$ column densities which differ by nearly a factor of
5.  To our knowledge, no laboratory measurements of the
$\lambda\lambda$1239,1240 $f$-values have been made.  Nor do other
theoretical calculations shed light on the issue, since they span an
even wider range than do the first two entries in Table 1 (Butler,
Mendoza, \& Zeippen 1984).
 
Magnesium is one of the most abundant metals and, because it readily
condenses into the solid form, is also likely to be one of the main
constituents of interstellar dust.  In addition, Mg provides a
diagnostic of electron density in the gas-phase of the ISM, through the
ionization ratio Mg$^0$/Mg$^+$.  An accurate assessment of the
importance of Mg in both the gas and dust clearly must start with
accurate column densities.  Until laboratory measurements are made, a
more definitive empirical determination of the $f$-values of these
important features seems the only way to clarify the role of Mg in the
the ISM and to exploit its diagnostic potential.  In this paper we
report the result of such an empirical determination, based on
absorption component analyses of the lines of sight towards the stars
$\xi$ Per (HD 24912), $\zeta$ Oph (HD 149757), and HD 93521.  The new
analyses and the resultant $f$-values are described in \S 2.  In \S 3,
the implications of the revised Mg$^+$ column densities for
interstellar gas and dust studies are discussed.
 
\section{New Analysis and Results}

The analysis used here is the ``component method,'' where complex
interstellar profiles are assumed to result from the superposition of
the absorption from a number of distinct interstellar clouds.  The goal
of the component method is to determine the set of velocity centroids,
velocity widths, and individual atomic column densities which
characterize each cloud along a line-of-sight.  This analysis has been
applied by us to a number of different sightlines, combining {\it GHRS}
UV data with ground-based optical data and, sometimes, 21-cm data.  The
most recent results are described by Fitzpatrick \& Spitzer (1997;
hereafter Paper IV) for the star HD 215733; see that paper (and the
references cited) for a complete description of the process.  The
analysis in this paper is the same as in the previous studies, except
that the Mg II $\lambda\lambda$1239,1240 $f$-values are taken as free
parameters to be determined from the fit to the data, but constrained
such that $f(\lambda1240) = \frac{1}{2}f(\lambda1239)$.  The basic data
processing was also carried out as described in the earlier papers,
including the normalization of the interstellar absorption profiles
using Legendre polynomials.
 
For this study, we applied the component analysis to data for three
sightlines, toward the stars $\zeta$ Oph, $\xi$ Per, and HD 93521.  The
final model for $\zeta$ Oph is based on 14 UV transitions from 7
species (including O$^0$, Mg$^+$, Si$^+$, Cr$^+$, Mn$^+$, Fe$^+$, and
Zn$^+$); for $\xi$ Per, on 18 UV transitions from 9 species (including
Mg$^0$, Mg$^+$, Si$^+$, S$^0$, S$^+$, Cr$^+$, Mn$^+$, Fe$^+$, and
Zn$^+$); and for HD 93521, on 20 UV and optical transitions from 9
species (including C$^{+*}$, Na$^0$, Mg$^0$, Mg$^+$, Si$^+$, S$^+$,
Ca$^+$, Mn$^+$, and Fe$^+$).  All four UV Mg$^+$ lines
($\lambda\lambda$1239,1240,2796,2803) were included in each of the
three analyses.  The UV data for $\zeta$ Oph and $\xi$ Per were taken
from the {\it HST} Archives and, with one exception, consist of {\it
GHRS} echelle observations (Ech-A and Ech-B) obtained with the stars
centered in the small science aperture (SSA).  The exception is the Mg
II $\lambda\lambda$2796,2803 observation toward $\zeta$ Oph, which was
performed using the G270M intermediate resolution grating and the SSA.
Previous analyses of these data can be found in Savage, Cardelli, \&
Sofia (1992); Savage et al. (1991); and Cardelli et al. (1991).  The
data for HD 93521 consist of new Cycle 5 {\it GHRS} observations
obtained in November 1996 for this study, our own earlier {\it GHRS}
data which were analyzed by Spitzer \& Fitzpatrick (1993; Paper I), and
new high resolution ground-based observations of Ca II $\lambda$3933
and Na I $\lambda\lambda$5890,5896.  All the {\it GHRS} observations of
HD 93521 were made using the echelle modes and the SSA.

The component models for $\zeta$ Oph and $\xi$ Per were based initially
on the Ca II models of Welty, Morton, \& Hobbs (1996), then iterated to
find the best fits to the UV data.  The HD 93521 model was based on our
earlier results in Paper I.  The details of these new models will be
discussed in future papers (Fitzpatrick \& Spitzer, in preparation).
Here we present only the results that pertain to the Mg$^+$ $f$-value
determination.  The normalized line profiles of the four UV Mg$^+$
lines are shown for each of the three stars in Figure 1 (filled
circles).  Note that the vertical scale is expanded for the HD 93521
$\lambda\lambda$1239,1240 lines and that the HD 93521 data were
obtained at a higher sampling frequency (4 data points per diode) than
those for $\zeta$ Oph and $\xi$ Per (2 points per diode).  The
theoretical profiles resulting from the three component analyses are
shown as the solid curves in Figure 1.  The $\lambda\lambda$1239,1240
$f$-values derived from these analyses are listed in Table 1.  These
results --- as can be seen from the associated 1-$\sigma$ errors, which
incorporate uncertainties in the component models, in the continuum
level determinations, and from random noise in the data --- are all
mutually consistent at the 1-$\sigma$ level and yield final weighted
mean values as given in the last line of Table 1.

The fitting procedure which produced the results in Figure 1 and Table
1 is a complex, nonlinear process for determining many variables
simultaneously.  The solution for the $f$-values can be viewed in
simplified form, however, as follows: {\it relative} column densities
for the various absorption components of Mg$^+$ are determined from the
weak $\lambda\lambda$1239,1240 lines, which show the component
structure clearly; the {\it total} Mg$^+$ column density and, hence,
the $\lambda\lambda$1239,1240 $f$-values are then found by scaling the
relative column densities to fit the profiles of the strong
$\lambda\lambda$2796,2803 lines.  The agreement of the results from
analyses of three very different sightlines is satisfying and we
believe the weighted means in Table 1 are the best currently available
estimates of the $\lambda\lambda$1239,1240 $f$-values.

\section{Discussion}

The new $f$-values in Table 1 require a revision of all published
Mg$^+$ column densities derived from the $\lambda\lambda$1239,1240
lines.  For those results based on the Hibbert et al. (1983) $f$-values
(e.g., Paper IV and earlier papers in that series), $N({\rm Mg}^+)$
should be scaled {\it downwards} by a factor of $\sim$2.4; for those
based on the Sofia et al. (1994) results (e.g., Sembach \& Savage 1996),
$N({\rm Mg}^+)$ should be scaled {\it upwards} by a factor of
$\sim$2.0.  These changes in $N({\rm Mg}^+)$ affect two distinct
analyses: the determination of $n_e$ in the ISM based on the
Mg$^0$/Mg$^+$ ratio and the composition of interstellar dust grains.
In this section we briefly consider these two areas.

Evidence was presented in Paper IV that very different values of $n_e$
can be obtained for the same interstellar cloud, depending on which
$n_e$ diagnostic is employed.  In particular, values of $n_e$ derived
from ionization equilibrium involving a neutral species (e.g.,
C$^0$/C$^+$, Mg$^0$/Mg$^+$, and S$^0$/S$^+$) were found to be
systematically larger (by 0.4 to 1.0 dex) than values derived from
Ca$^+$/Ca$^{++}$ ionization equilibrium or from collisional excitation
of fine structure levels in C$^+$.  (These latter two diagnostics
appear to be in accord.)  The revised $f$-values presented here have
the effect of increasing the values of $n_e$ derived in Paper IV from
Mg$^0$/Mg$^+$ by $\sim$0.4 dex.  From Figure 6 of that paper it can be
seen that this enhances the discrepancy between these values and the
Ca$^+$ results --- but also decreases the scatter among the neutral
species.  $N_e$ values from the neutrals now appear consistent with
each other to within $\sim$0.2 dex, but are systematically greater than
those from other diagnostics by a factor of 0.8 to 1.0 dex.  The cause
of this discrepancy is not known, although it seems likely that
unaccounted for processes are involved in the recombination of at least
some singly-ionized species.

Measurements of $N({\rm Mg}^+)$ in H I gas, when coupled with an
assumed intrinsic abundance of Mg, indicate how much Mg is condensed
into interstellar dust grains.  Paper I showed, using the Hibbert et
al.  (1983) $f$-values and assuming solar system composition, that the
total numbers of Mg + Fe atoms incorporated into dust at any general
depletion level is about twice the number of Si atoms in the dust.
This would be consistent with --- but does not require --- that all the
depleted Fe, Mg, and Si are located in silicate grains.  In contrast,
Sofia et al. (1994) found, using their much larger estimate of the
$\lambda\lambda$1239,1240 $f$-values, that the ratio of Mg + Fe atoms
to Si atoms in dust grain ``cores'' is much greater than can arise from
silicate grains, indicating that the equivalent of 50$\%$ or more of
{\it both} the depleted Fe and Mg atoms must reside in another dust
grain population, perhaps oxides.  The new $\lambda\lambda$1239,1240
$f$-values allow these contrasting conclusions about dust grain
composition to be reconciled.

Figure 2 compares the ``relative gas-phase abundances'' (or
``depletions;'' see Eq. 1 in Paper IV) of Mg and Si for those
sightlines with the most secure Mg$^+$ column densities.  The key to
the symbols is shown on the figure and the sources of the data given in
the legend.  Unless otherwise indicated, uncertainties are comparable
to or less than the symbol sizes.  The dotted line shows a linear fit
to the data and has a slope of 0.82 and an intercept of $-0.17$ dex.
Future results may show such a simple linear relation to be inadequate,
but at the current time it represents the trend in the data reasonably
well.  Along this line, from [Si/H] = --0.2 to [Si/H] = --1.5, the
ratio of Mg atoms to Si atoms in the gas varies relatively little, only over
the range 0.8 to 1.3.

The results in Figure 2, combined with comparable data for Fe (shown in
Figure 2 of Fitzpatrick 1996), yield limits on the incorporation of Mg
and Fe in silicate grains.  In this particular case, and in many
others, the strongest constraints on dust grain composition come from
the sightlines with the smallest measured amounts of depletion.  Along
such sightlines, we have [Si/H] $\simeq$ --0.2, [Mg/H] $\simeq$ --0.35,
and [Fe/H] $\simeq$ --0.55, which imply 37\%, 55\%, and 72\% of the Si,
Mg, and Fe in dust grains, respectively, using the solar system
composition as the abundance reference.  These yield total dust-phase
ratios of Mg/Si = 1.6 and Fe/Si = 1.8.  We thus conclude, in
qualitative agreement with Sofia et al. (1994), that the total ratio of
Fe + Mg atoms to Si atoms in the dust (3.4:1) is greater than the
maximum ratio that can be accounted for by silicate grains alone (i.e.,
2:1).  If the solar system composition is appropriate for the ISM, then
these results imply that at least 50\% of the total available Mg (i.e.,
gas + dust) {\it or\/} 50\% of the total Fe (or some comparable
combination, such as 25\% of each) must reside in some grain population
other than silicates.  Alternatively, the Mg, Fe, and Si depletions
would be completely consistent with silicates grains if the intrinsic
interstellar abundances of Mg or Fe, or both, were reduced relative to
Si by these same amounts.

As noted by Fitzpatrick (1996), the gas-phase abundances of some
elements, notably Fe, always appear to be subsolar even along the most
lightly depleted sightlines.  This result can be interpreted as
evidence either for a population of extremely hardy dust grains or for
intrinsically subsolar ISM abundances for these elements.  Even with
this ambiguity in interpretation, however, the observations of the
smallest depletions still serve to provide lower limits to the ISM
abundances.  Magnesium, like Fe and Si, has not yet been detected
reliably with a depletion approaching zero.  The smallest well-measured
Mg depletions are $\sim$--0.3 dex (see Figure 2) and indicate that the
ISM abundance of Mg is at least 50\% of the solar system value.  We
note that Jenkins \& Wallerstein (1996) find essentially zero depletion
for Si and Mg towards the halo star HD 120086, but the column densities
for Mg$^+$, Si$^+$, and the normalizing species S$^+$ have
uncertainties of several tenths of a dex.  Followup observations of
this sightline could be important for discriminating between the two interpretations of the minimum depletions.

In summary, we have presented new, precise empirical measurements of
the $f$-values of the important Mg II $\lambda\lambda$1239,1240 lines
based on component modeling of absorption along three lines of sight.
These new values require factor-of-2 changes (both increases and
decreases) in published Mg$^+$ column densities, which affect
measurements of interstellar electron densities and determinations of
the composition of interstellar gas and dust.  These data confirm that
not all the Fe and Mg atoms ``missing'' from the gas phase of the ISM
can be incorporated in silicate grains.  With the completion of this
analysis, the set of $f$-values available for the most important
diagnostic species in the diffuse, neutral ISM (e.g, Mg$^+$, Si$^+$,
S$^+$, Fe$^+$, and Zn$^+$) appears to be extremely well-determined, and
finally justifies the common practice of ignoring $f$-value
uncertainties when quoting column density uncertainties!

%
%

\acknowledgements

I am grateful to the referee U.J. Sofia for helpful comments.  This
research has been supported by Space Telescope Science Institute Grant
GO-05878.01-94A to Princeton University and by Rice University
subcontract SC-437-5-16591 with Princeton University, supported in turn
by NASA grant NAG5-1626 to Rice.

%
%


%
%

\newpage
\begin{deluxetable}{lcc}
\tablenum{1}
\tablecaption{Oscillator Strengths for Mg II $\lambda\lambda1239,1240$}
\tablewidth{30pc}
\tablehead{
\colhead{Source}                 &
\colhead{$f$($\lambda$1239.9)}   &
\colhead{$f$($\lambda$1240.4)}   \\
\colhead{(1)}                    &
\colhead{(2)}                    &
\colhead{(3)} }
 
\startdata
Hibbert et al. 1983\tablenotemark{a} .......... & $2.7\pm0.7(-4)$ & $1.3\pm0.3(-4)$  \nl
Sofia et al. 1994\tablenotemark{b} ..............   & $12.5\pm\hspace{0.001em}^{7.8}_{3.8}(-4)$ &  $6.25\pm\hspace{0.001em}^{3.85}_{1.92}(-4)$   \nl

\nl
\multicolumn{3}{c}{This Paper} \nl
\nl

HD 93521 ........................ & $5.65\pm1.09(-4)$  & $2.82\pm0.55(-4)$  \nl
$\xi$ Per ............................... & $6.86\pm0.61(-4)$  & $3.43\pm0.30(-4)$  \nl
$\zeta$ Oph .............................. & $6.10\pm0.60(-4)$  & $3.05\pm0.30(-4)$  \nl
Weighted Mean................ & ${\bf6.4\pm0.4(-4)}$  & ${\bf3.2\pm0.2(-4)}$  \nl
\enddata
\tablecomments{In cols. (2) and (3), the notation $a(-b)$ means $a\times10^{-b}$}.
\tablenotetext{a}{Theoretical calculation.}
\tablenotetext{b}{Empirical result, based on $\xi$ Per data.}
\end{deluxetable}

%
%

\clearpage
\begin{figure}
\figurenum{1}
\caption{Normalized line profiles of the interstellar Mg II
$\lambda\lambda$1239, 1240, 2796, and 2803 lines are shown for the
three stars $\zeta$ Oph, $\xi$ Per, and HD 93521 (filled circles).  All
data were obtained with the {\it GHRS} using the small science aperture
and the echelle modes, except the $\lambda\lambda$2796, 2803
observations for $\zeta$ Oph, which were made with the G270M
intermediate resolution grating.  The data for HD 93521 have a sampling
frequency of 4 data points per {\it GHRS} diode, compared to 2 points
per diode for the other stars.  The solid curves through the data
points are the theoretical profiles resulting from the component
analyses.   Note that the velocity scales are different for the three
stars and that the $\lambda\lambda$1239, 1240 lines for HD 93521 are
shown with an expanded vertical scale.}

\notetoeditor{NOTE TO EDITOR: Figure 1 is intended to run horizontally
across the top half of one page}
\end{figure}

\begin{figure}
\figurenum{2}
\caption{Relative gas-phase abundances (or ``depletions'') of Mg
plotted against those of Si for the lines of sight toward the halo
stars HD 18100, HD 93521, and HD 215733 and toward the disk stars HD
68273, $\xi$ Per, and $\zeta$ Oph (see equation 1 of Paper IV).  Small
symbols indicate results for individual absorption components; large
symbols indicate integrations over multiple components as detailed
below.  The dashed line shows a linear fit to the data, with a slope
and intercept of 0.82 and --0.17, respectively.  The gas-phase
abundances for HD 18100, HD 68273, HD 93521, and HD 215733 were
normalized using the observed S$^+$ column densities, for $\xi$ Per by
the total H column density (H$^0$ + H$_2$), and for $\zeta$ Oph by the
observed Zn$^+$ column densities, assuming that [Zn/H]$_g$ = --0.29 for
all the components.  Column density errors in the normalization species
(which are probably less than 0.1 dex) move the data points along a
line of slope = 1, introducing negligible scatter in the observed
relation.  Results for HD 18100 and $\xi$ Per are integrated over all
absorption components; for HD 93521 results are show for integrations
over the velocity ranges $v <$  $-30$ km s$^{-1}$ and $v >$  $0$ km
s$^{-1}$; and for HD 215733 for $v <$ $-40$ km s$^{-1}$ and $v >$ $-40$
km s$^{-1}$.  Data for HD 93521, $\xi$ Per, and $\zeta$ Oph are from
this paper; for HD 18100 from Savage \& Sembach 1996b; For HD 68273
from Fitzpatrick \& Spitzer 1994; and for HD 215733 from Paper IV.}

\notetoeditor{NOTE TO EDITOR: Figure 2 is intended to occupy 1 column.}
\end{figure}

%
%

\begin{figure}
\figurenum{1}
\plotone{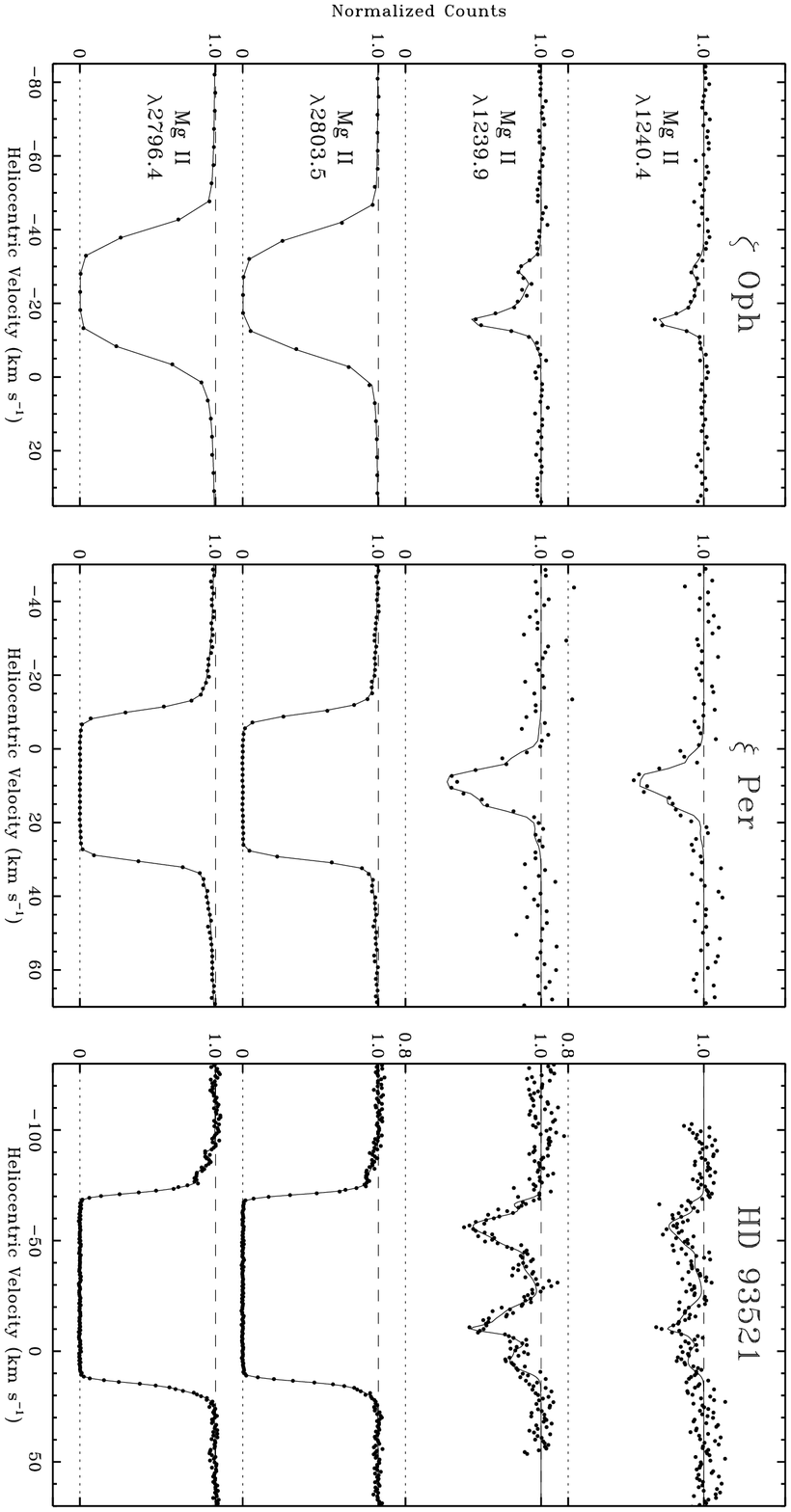}
\caption{}
\end{figure}

\begin{figure}
\figurenum{2}
\plotone{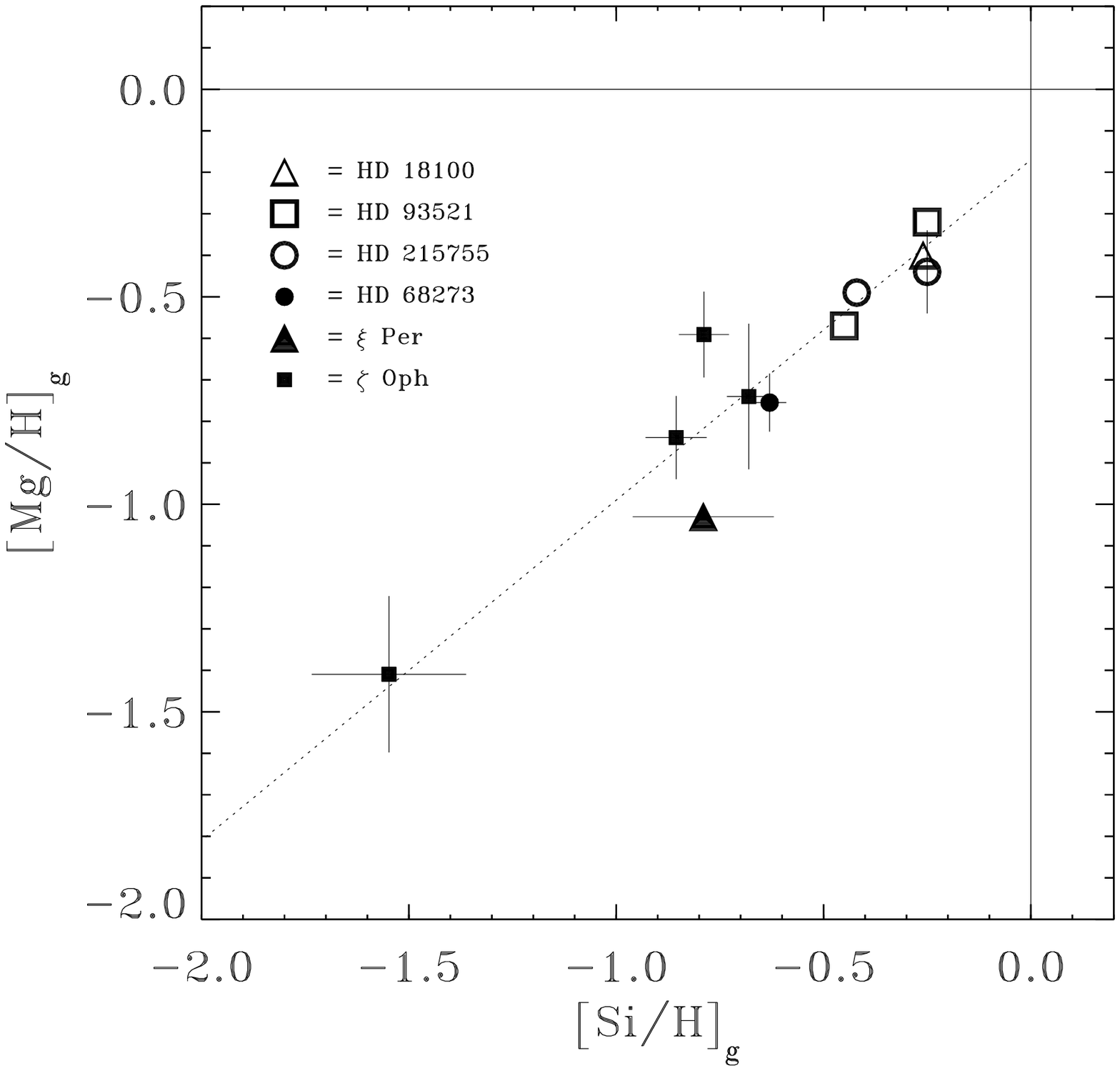}
\caption{}
\end{figure}

\end{document}